\documentclass[12pt]{article}

\newcommand{\be}{\begin{equation}}
\newcommand{\ee}{\end{equation}}
\newcommand{\bi}[1]{\bibitem{#1}}

\begin{document}
\begin{center}
{\it Physics Letters A 372 (2006) 2984-2988}
\end{center}
\vskip 5 mm

\begin{center}
{\Large \bf Fractional Heisenberg Equation}
\vskip 5 mm

{\large \bf Vasily E. Tarasov}\\
\vskip 3mm

{\it Skobeltsyn Institute of Nuclear Physics, \\
Moscow State University, Moscow 119991, Russia \\
E-mail: tarasov@theory.sinp.msu.ru }

\end{center}

\begin{abstract}
Fractional derivative can be defined as a fractional power of derivative.
The commutator $(i/\hbar)[H, \ . \ ]$, 
which is used in the Heisenberg equation, 
is a derivation on a set of observables. 
A derivation is a map that satisfies the Leibnitz rule.
In this paper, we consider a fractional derivative on
a set of quantum observables as a fractional power of the
commutator $(i/\hbar)[H, \ . \ ]$.
As a result, we obtain a fractional generalization of the
Heisenberg equation.  
The fractional Heisenberg equation is exactly solved 
for the Hamiltonians of free particle and harmonic oscillator.
The suggested Heisenberg equation generalize a notion of 
quantum Hamiltonian systems to describe 
quantum dissipative processes.
\end{abstract}





\newpage
\section{Introduction}

The fractional calculus has a long history from 1695, 
when the derivative of order $1/2$ has been described 
by Leibniz \cite{OS,SKM}.
The theory of derivatives and integrals of non-integer order 
goes back to Leibniz, Liouville, Grunwald, Letnikov and Riemann.
There are many books about fractional calculus and 
fractional differential equations \cite{OS,SKM,Podlubny,KST}.
Derivatives of fractional order, and 
fractional differential equations 
have found many applications in recent studies in physics
(see for example \cite{Zaslavsky1,GM,Zaslavsky2,MS} 
and references therein).

In the quantum kinematics, 
the observables are given by self-adjoint operators.
The dynamical description of system is given by a superoperator, 
which is a rule that assigns to each operator exactly  one operator.
The natural description of the motion is in terms of the 
infinitesimal change of the system.
The equation of motion for quantum observable
is called the Heisenberg equation. 
For Hamiltonian systems, the infinitesimal superoperator 
of the Heisenberg equation
is defined by some form of derivation.

Fractional derivative can be defined as a fractional power
of derivative (see for example \cite{IJM}).
It is known that the infinitesimal generator $(i/\hbar)[H, \ . \ ]$,
which is used in the Heisenberg equation, 
is a derivation of observables.
A derivation is a linear map $D$, which satisfies the Leibnitz rule
$D(AB)=(DA)B+ A(DB)$ for all operators $A$ and $B$.
In this paper, we consider a fractional derivative on
a set of observables as a fractional power of derivative.
As a result, we obtain a fractional generalization of the
Heisenberg equation.
It allows us to generalize a notion of quantum Hamiltonian systems. 
Note that fractional generalization of 
classical Hamiltonian systems has been suggested in \cite{FracHam}.
The suggested fractional Heisenberg equation is exactly solved 
for the Hamiltonians of free particle and harmonic oscillator.
A quantum system that is presented by fractional Heisenberg equation 
can be considered as a dissipative system \cite{Weiss}.
Fractional derivatives can be used as a possible approach to 
describe an interaction between the system and an environment.
Note that it is possible to consider quantum dynamics with 
low-level fractionality by some generalization of 
method suggested in \cite{TZ2} (see also \cite{TP}).

In Section 2, the fractional power of derivative and 
the fractional Heisenberg (FH) equation are suggested.
In Section 3, the properties of time evolution, 
which is described by the fractional equation, are considered.
In Section 4, the FH equation for free particle is solved.
In Section 5. the solution of FH equation with harmonic 
oscillator Hamiltonian is derived.

\newpage
\section{Fractional derivative and Heisenberg equation}

Let us consider a set of quantum observables in the Heisenberg picture. 
A superoperator ${\cal L}$ 
is a rule that assigns to each operator $A$ 
exactly one operator ${\cal L}(A)$. 
(About superoperator formalism see for example \cite{Quant1,Quant2,Quant3}
and references therein.)  For Hamiltonian $H$, 
let $L^{-}_H$ be the superoperator given by
\[ L^{-}_H A=\frac{1}{i\hbar} (HA-AH) . \]
The operator differential equation
\be \label{Heis1} \frac{d}{dt} A_t=-L^{-}_H A_t \ee
is called the Heisenberg equation for Hamiltonian systems.
The time evolution of a Hamiltonian system is 
induced by the Hamiltonian $H$.

It is interesting to obtain a fractional generalization
of equation (\ref{Heis1}).
We will consider here concept of fractional power for $L^{-}_H$.
If $L^{-}_H$ is a closed linear superoperator
with an everywhere dense domain $D(L^{-}_H)$,
having a resolvent $R(z,L^{-}_H)=(zL_I-L^{-}_H)^{-1}$ 
on the negative half-axis, 
then there exists \cite{Bal,Yosida,Krein} the superoperator
\be \label{LaA}
-(L^{-}_H)^{\alpha}=\frac{\sin \pi \alpha }{\pi} 
\int^{\infty}_0 dz\, 
z^{\alpha-1} R(-z,L^{-}_H) \, L^{-}_H
\ee
defined on $D(L^{-}_H)$ for $0< \alpha <1$. 
The superoperator $(L^{-}_H)^{\alpha}$ is 
a {\it fractional power of the Lie left superoperator}.

As a result, we obtain the equation
\be \label{Heis2} 
\frac{d}{dt} A_t=-(L^{-}_H)^{\alpha} A_t , \ee
where $t$  and $H/\hbar$ are dimensionless variables.
This is the {\it fractional Heisenberg equation}.

Note that this equation cannot be presented in the form
\be 
\frac{d}{dt} A_t=-L^{-}_{H_{new}} A_t=\frac{i}{\hbar} [H_{new}, A_t] 
\ee
with some operator $H_{new}$.
Therefore, quantum systems described by (\ref{Heis2}) 
are not Hamiltonian systems.
The systems will be called the fractional Hamiltonian systems (FHS). 
Usual Hamiltonian systems can be considered as 
a special case of FHS. 

If we consider the Cauchy problem for equation (\ref{Heis1}) 
in which the initial condition is given at the time $t=0$ by $A_0$,
then its solution can be written in the form
\[ A_t=\Phi_t A_0. \]
The one-parameter superoperators $\Phi_t$, $t\ge 0$ 
have the properties
\[ \Phi_t \Phi_s=\Phi_{t+s}, \quad (t,s >0) , \quad \Phi_0=L_I , \]
where $L_I$ is unit superoperator ($L_I A=A$).
Then the set of $\Phi_t$, $t\ge 0$, is called the semi-group.
Then the superoperator $L^{-}_H$ is called the generating superoperator,
or infinitesimal generator, of the semi-group $\{\Phi_t, t\ge 0\}$.

Let us consider the Cauchy problem for 
fractional Heisenberg equation (\ref{Heis2}) 
in which the initial condition is given by $A_0$.
Then its solution can be presented in the form
\[ A_t(\alpha)=\Phi^{(\alpha)}_t A_0, \]
where the superoperators $\Phi^{(\alpha)}_t$, $t>0$, 
form a semi-group which will be called the {\it fractional semi-group}.
The superoperator $(L^{-}_H)^{\alpha}$ is infinitesimal generator of 
the semi-group $\{\Phi^{(\alpha)}_t, t\ge 0\}$.

\section{Properties of fractional time evolution}

Let us consider some properties of time evolution described by
a fractional semi-group $\{\Phi^{(\alpha)}_t, t\ge 0\}$. \\

(1) The superoperators $\Phi^{(\alpha)}_t$ can be constructed
in terms of the superoperators $\Phi_t$
by the Bochner-Phillips formula \cite{Bochner,Phillips,Yosida}
\be \label{BPf2}
\Phi^{(\alpha)}_t=
\int^{\infty}_0 ds f_{\alpha}(t,s) \Phi_s , 
\quad (t>0) , \ee 
where $f_{\alpha}(t,s)$ is defined by
\be \label{fats}
f_{\alpha}(t,s)=\frac{1}{2\pi i} \int^{a+\i\infty}_{a-i\infty}
dz \, \exp (sz-tz^{\alpha}) ,
\ee
where $a,t>0$, $s \ge 0 $, and $0<\alpha <1$.
The branch of $z^{\alpha}$ is so taken that $Re(z^{\alpha})>0$
for $Re(z)>0$.
This branch is a one-valued function in the $z$-plane 
cut along the negative real axis.
The convergence of this integral is obviously
in virtue of the convergence factor $\exp(-tz^{\alpha})$.
By denoting the path of integration in (\ref{fats})
to the union of two paths $r\, \exp(-i \theta)$, and 
$r\, \exp(+i \theta)$, where $r \in (0,\infty)$, 
and $\pi/2 \le \theta \le \pi$, we obtain
\[ f_{\alpha}(t,s)=\frac{1}{\pi} \int^{\infty}_0  dr \,
\exp (sr \cos \theta - t r^{\alpha} \cos (\alpha \theta)) \cdot \]
\be \label{freal}
 \cdot \sin (sr \sin \theta - 
t r^{\alpha} \sin (\alpha \theta)+ \theta) . \ee
If we have a solution $A_t$ of the Heisenberg equation (\ref{Heis1}),
then formula (\ref{BPf2}) gives the solution
\be 
A_t(\alpha)=\int^{\infty}_0 ds f_{\alpha}(t,s) A_s , 
\quad (t>0) \ee
of fractional equation (\ref{Heis2}). \\

(2) In quantum theory, the most important is the class of 
real superoperators. Let $A^{*}$ be an adjoint of $A$.
A quantum observable is a self-adjoint operator. 
If $\Phi_t$ is a real superoperator
and $A$ is a self-adjoint operator $A^{*}=A$, 
then the operator $A_t=\Phi_t A$ is self-adjoint, i.e., 
\[ (\Phi_t A)^{*}=\Phi_t A . \]
A superoperator as a map from a set of observables into 
itself should be real. All possible dynamics, i.e., 
temporal evolutions of quantum observables,
should be described by real superoperators. 
Therefore the following statement is very important.
If $\Phi_t$ is a real superoperator, 
then $\Phi^{(\alpha)}_t$ is real. 
It can be proved by using the Bochner-Phillips formula
and equation (\ref{freal}).  \\

(3) An adjoint superoperator of $\Phi_t$ 
is a superoperator $\bar \Phi_t$, such that 
\[ Tr [(\bar \Phi_t A)^{*} B] =Tr [A^{*} \Phi_t (B)] . \]
Let us give the basic statement regarding 
the adjoint superoperator. 
If $\bar \Phi_t$ is an adjoint superoperator of $\Phi_t$, 
then the superoperator
\[ \bar \Phi^{(\alpha)}_t =
\int^{\infty}_0 ds f_{\alpha}(t,s) \bar \Phi_s , 
\quad (t>0) , \]
is an adjoint superoperator of $\Phi^{(\alpha)}_t$.
We prove this statement by using the Bochner-Phillips formula:
\[ Tr [(\bar \Phi^{(\alpha)}_t A)^{*} B] =
\int^{\infty}_0 ds f_{\alpha}(t,s) 
Tr [(\bar \Phi_s A)^{*} B] = \]
\[ =\int^{\infty}_0 ds f_{\alpha}(t,s) 
Tr [A^{*} \Phi_s (B)] = Tr [A^{*} \Phi^{(\alpha)}_t (B)] . \]

Let $\{\bar \Phi_t, t>0\}$ be a semi-group, 
such that the density matrix operator $\rho_t=\bar \Phi_t \rho_0$
is described by the von Neumann equation
\[ \frac{d}{dt} \rho_t=\frac{1}{i\hbar}[H,\rho_t] .  \]
Then the semi-group $\{\bar \Phi^{(\alpha)}_t, t>0\}$
describes the evolution of the density operator
\[ \rho_t(\alpha)=\bar \Phi^{(\alpha)}_t \rho_0 \]
by the fractional equation
\[ \frac{d}{dt} \rho_t=-(-L^{-}_H)^{\alpha} \rho_t . \]
This is the {\it fractional von Neumann equation}. \\

(4) Let $\Phi_t$, $t>0$, be a positive one-parameter 
superoperator, i.e., $\Phi_t A \ge 0$ for $A \ge 0$. 
Using the Bochner-Phillips formula and the property
$f_{\alpha}(t,s) \ge 0$, $(s>0)$, it is easy to prove 
that the superoperators $\Phi^{(\alpha)}_t$
are also positive, i.e., $\Phi^{(\alpha)}_t A \ge 0$ for $A \ge 0$. \\

(5) It is known that $\bar \Phi_t$ is a real superoperator if
$\Phi_t$ is real. 
Analogously, if $\Phi^{(\alpha)}_t$ is a real superoperator, then
$\bar \Phi^{(\alpha)}_t$ is real.


\section{Fractional free particle}

Let us consider the Hamiltonian $H=P^2/2m$,
where $P$ is dimensionless variable,
and $m^{-1}$ has the action dimension.
Then the Heisenberg equation (\ref{Heis1}) describes a free 
one-dimensional particle.
For $A=Q$, and $A=P$, equation (\ref{Heis1}) gives
\[ \frac{d}{dt} Q_t=\frac{1}{m} P_t, \quad 
\frac{d}{dt} P_t=0 . \]
The well-known solutions of these equations are
\be \label{Hsol1}
Q_t=Q_0 +\frac{t}{m} P_0 , \quad P_t=P_0 . \ee
Using these solutions and the Bochner-Phillips formula, 
we will obtain the solutions of fractional Heisenberg equations
\be \label{ex1}
\frac{d}{dt} Q_t=- \frac{1}{m^{\alpha}} (L^{-}_{P^2})^{\alpha} Q_t, 
\quad \frac{d}{dt} P_t=0 . \ee
Note that $(L^{-}_{P^2})^{\alpha} \ne L^{-}_{P^{2\alpha}}$. 
The solutions of fractional equations (\ref{ex1}) have the forms
\[ Q_t(\alpha)=\Phi^{(\alpha)}_t Q_0=
\int^{\infty}_0 ds f_{\alpha}(t,s) Q_s , 
\quad P_t(\alpha)=P_0 , \]
where $Q_s$ is given by (\ref{Hsol1}).
Then
\[ Q_t=Q_0 +\frac{1}{m} g_{\alpha}(t) P_0 , \quad P_t=P_0 , \] 
where
\[ g_{\alpha}(t)= \int^{\infty}_0 ds f_{\alpha}(t,s) \, s . \]

If $\alpha=1/2$, then we have
\[ g_{1/2}(t)= \frac{t}{2 \sqrt{\pi}} \int^{\infty}_0 ds \, 
\frac{1}{\sqrt{s}} e^{-t^2/4s} = \frac{t^2}{2} . \]
Then
\be \label{free}
Q_t=Q_0 -\frac{t^2}{2m} P_0 , \quad P_t=P_0 . \ee
These equations describe a fractional free motion for $\alpha=1/2$.
For the operators (\ref{free}), 
the average values and dispersions have the form
\[ <Q_t>=x_0 -\frac{t^2}{2m} p_0 , \quad <P_t>=p_0 , \] 
and
\[ D_P(t)=\frac{\hbar^2}{2b^2} , \quad 
D_Q(t)=\frac{b^2}{2} \Bigl(1+\frac{\hbar^2 t^4}{m^2 b^4} \Bigr) .\]
Here we use the coordinate representation 
and the pure state
\be \label{Psi0}
\Psi(x)=<x|\Psi>= (b\sqrt{\pi})^{-1/2} \exp\Bigl(-\frac{(x-x_0)^2}{2b} +
\frac{i}{\hbar} p_0x \Bigr) . \ee
The average value and dispersion are defined by the well-known equations
\[ <A_t>=Tr[|\Psi><\Psi|A_t] =<\Psi|A_t|\Psi> ,  \]
\[ D_A(t)=<A^2_t>-<A_t>^2=<\Psi|A^2_t|\Psi>-<\Psi|A_t|\Psi>^2 . \]

\section{Fractional Heisenberg equation for harmonic oscillator}

Let us consider the Hamiltonian 
\be \label{oscHam}
H=\frac{1}{2m} P^2 +\frac{m\omega^2}{2} Q^2, \ee
where $t$ and $P$ are dimensionless variables.
Then equation (\ref{Heis1}) describes a harmonic oscillator.
For $A=Q$, and $A=P$, equation (\ref{Heis1}) gives
\[ \frac{d}{dt} Q_t=\frac{1}{m} P_t, \quad 
\frac{d}{dt} P_t=-m \omega^2 Q_t . \]
The well-known solutions of these equations are
\[ Q_t=Q_0 \cos (\omega t) +\frac{1}{m \omega} P_0 \sin (\omega t) , \]
\be \label{osc1}
P_t=P_0 \cos (\omega t) - m \omega Q_0 \sin (\omega t) . \ee 
Using these solutions and the Bochner-Phillips formula, 
we will obtain the solutions of fractional Heisenberg equations
\be \label{ex2}
\frac{d}{dt} Q_t=- (L^{-}_{H})^{\alpha} Q_t , \quad 
\frac{d}{dt} P_t=- (L^{-}_{H})^{\alpha} P_t , \ee
where $H$ is defined by (\ref{oscHam}).
The solutions of fractional equations (\ref{ex2}) have the forms
\[ Q_t(\alpha)=\Phi^{(\alpha)}_t Q_0=
\int^{\infty}_0 ds f_{\alpha}(t,s) Q_s , \]
\be \label{osc2} 
P_t(\alpha)=\Phi^{(\alpha)}_t P_0=
\int^{\infty}_0 ds f_{\alpha}(t,s) P_s . \ee
Substitution of (\ref{osc1}) into (\ref{osc2}) gives
\be \label{Hsol2a}
Q_t=Q_0 C_{\alpha}(t) +\frac{1}{m \omega} P_0 S_{\alpha}(t) , \ee
\be \label{Hsol2b}
P_t=P_0 C_{\alpha}(t) - m \omega Q_0 S_{\alpha}(t) , \ee
where
\[ C_{\alpha}(t)=\int^{\infty}_0 ds \, f_{\alpha}(t,s)\, \cos(\omega s) , \]
\[ S_{\alpha}(t)=\int^{\infty}_0 ds \, f_{\alpha}(t,s)\, \sin(\omega s) . \]
Equations (\ref{Hsol2a}) and (\ref{Hsol2b}) describe 
solutions of the fractional Heisenberg equations (\ref{ex2})
for quantum harmonic oscillator. 

If $\alpha=1/2$, then 
\[ C_{1/2}(t)=\frac{t}{2 \sqrt{\pi}} \int^{\infty}_0 ds \, 
\frac{\cos(\omega s)}{s^{3/2}} \, e^{-t^2/4s} , \]
\[ S_{1/2}(t)=\frac{t}{2 \sqrt{\pi}} \int^{\infty}_0 ds \, 
\frac{\sin(\omega s)}{s^{3/2}} \, e^{-t^2/4s} . \]
These functions can be presented through the Macdonald function
(see \cite{Prudnikov}, Sec. 2.5.37.1.) such that
\[ C_{1/2}(t)= \Bigl(\frac{\omega t^2}{4 \pi}\Bigr)^{1/4} 
\Bigl[ e^{+ \pi i/8} K_{-1/2} \Bigl(2 e^{+\pi i/4} 
\sqrt{ \frac{\omega t^2}{4} } \Bigr)+
e^{-\pi i/8} K_{-1/2} \Bigl(2 e^{-\pi i/4} 
\sqrt{ \frac{\omega t^2}{4} } \Bigr) \Bigr] , \]
\be \label{Bessel} S_{1/2}(t)= i \Bigl(\frac{\omega t^2}{4 \pi}\Bigr)^{1/4} 
\Bigl[ e^{+ \pi i/8} K_{-1/2} \Bigl(2 e^{+\pi i/4} 
\sqrt{ \frac{\omega t^2}{4} }\Bigr)-
e^{-\pi i/8} K_{-1/2} \Bigl(2 e^{-\pi i/4} 
\sqrt{ \frac{\omega t^2}{4} } \Bigr) \Bigr] , \ee
where $\omega>0$, and $K_{\alpha}(z)$ is the Macdonald function 
\cite{OS,SKM}, which is also called the modified Bessel function 
of the third kind.

Using (\ref{Psi0}), we get the average values 
\[ <Q_t>=x_0 C_{\alpha}(t) +\frac{1}{m \omega} p_0 S_{\alpha}(t) , \]
\[ <P_t>=p_0 C_{\alpha}(t) - m \omega x_0 S_{\alpha}(t) , \]
and the dispersions 
\[ D_P(t)=\frac{\hbar^2}{2b^2}
C^2_{\alpha}(t) +\frac{b^2 m^2 \omega^2}{2} S^2_{\alpha}(t) , \]
\[ D_Q(t)=\frac{b^2}{2}
C^2_{\alpha}(t) + \frac{\hbar^2}{2b^2 m^2 \omega^2} S^2_{\alpha}(t) . \]
As a result, the fractional harmonic oscillator is a simple dissipative systems. 
The dispersion of the wave packet is defined by these equations. 
The solutions are characterized by the fractional damping of 
the fractional harmonic oscillator.
The dumping is described by the modified Bessel function (\ref{Bessel})
of the third kind.
An important property of the evolution described by the fractional 
equations are that the solutions have power-like tails.

\section{Conclusion}

In this paper, we consider derivatives of noninteger order 
as fractional powers of derivative.
We derive a fractional generalization of the Heisenberg equation and 
a generalization of quantum Hamiltonian system to describe open quantum systems. 
A quantum system that is presented by fractional Heisenberg equation 
can be considered as a dissipative system.
Fractional derivatives can be used as an approach to 
describe an interaction between the quantum system and an environment.
This interpretation caused by following reasons.
Using the properties
\[ \int^{\infty}_0 f_{\alpha}(t,s)=1 , \quad \quad f_{\alpha}(t,s) \ge 0 \quad 
(for \ all \ \ s>0) , \]
we can assume that $f_{\alpha}(t,s)$ is a density of probability distribution.
Then the Bochner-Phillips formula (\ref{BPf2}) can be considered as
a smoothing of Hamiltonian evolution $\Phi_t$ with respect to time $s>0$.
This smoothing can be considered as an influence of 
the environment on the system. 
As a result, the parameter alpha can be used as a simple model 
of the interaction between the system and the environment.
Note that the quantum Markovian equations, 
which is also called the Lindblad equations \cite{Lindblad},
can be generalized by suggested approach 
to describe completely positive evolution 
of dissipative and open quantum systems.

\newpage

\end{document}